\documentclass[12pt,preprint]{aastex}

\newcommand{\bdv}[1]{\mbox{\boldmath$#1$}}

\def\au{{\rm AU}}

\def\masyr{{\rm mas}\,{\rm yr}^{-1}}
\def\kpc{{\rm kpc}}
\def\mas{{\rm mas}}
\def\muas{\mu{\rm as}}

\def\rel{{\rm rel}}

\def\geo{{\rm geo}}
\def\e{{\rm E}}
\def\bpi{{\bdv\pi}}
\def\bmu{{\bdv\mu}}

\def\btheta{{\bdv\theta}}

\begin{document}
\title{{\it Spitzer} Microlens Measurement of a Massive Remnant in
a Well-Separated Binary}

\author{
Y. Shvartzvald\altaffilmark{1,a},
A. Udalski\altaffilmark{2},
A. Gould\altaffilmark{3},
C. Han\altaffilmark{4},
V. Bozza\altaffilmark{5,6},
M. Friedmann\altaffilmark{7},
M.~Hundertmark\altaffilmark{8},\\
and\\
C. Beichman\altaffilmark{9},
G. Bryden\altaffilmark{1},
S. Calchi Novati\altaffilmark{9,5,10,b},
S. Carey\altaffilmark{11},
M. Fausnaugh\altaffilmark{3},
B.~S.~Gaudi\altaffilmark{3},
C. B. Henderson\altaffilmark{1,3,a},
T. Kerr\altaffilmark{12},
R. W. Pogge\altaffilmark{3},
W. Varricatt\altaffilmark{12},
B. Wibking\altaffilmark{3},
J. C.\ Yee\altaffilmark{13,c},
W. Zhu\altaffilmark{3},\\
(Spitzer team)\\
and\\
R. Poleski\altaffilmark{3},
M. Pawlak\altaffilmark{2},
M.\,K. Szyma{\'n}ski\altaffilmark{2},
J. Skowron\altaffilmark{2}
P. Mr{\'o}z\altaffilmark{2},
S. Koz{\l}owski\altaffilmark{2},
{\L}.~Wyrzykowski\altaffilmark{2},
P. Pietrukowicz\altaffilmark{2},
G. Pietrzy{\'n}ski\altaffilmark{2},
I. Soszy{\'n}ski\altaffilmark{2},
K. Ulaczyk\altaffilmark{14},\\
(OGLE group)\\
and\\
J.-Y.Choi\altaffilmark{4},
H. Park\altaffilmark{4},
Y. K. Jung\altaffilmark{4},
I.-G. Shin\altaffilmark{4},
M. D. Albrow\altaffilmark{15},
B.-G. Park\altaffilmark{16},
S.-L. Kim\altaffilmark{16},
C.-U. Lee\altaffilmark{16},
S.-M. Cha\altaffilmark{16,17},
D.-J. Kim\altaffilmark{16,17},
Y. Lee\altaffilmark{16,17},\\
(KMTNet group)\\
and\\
D. Maoz\altaffilmark{7},
S. Kaspi\altaffilmark{7},\\
(Wise group)\\
and\\
R. A. Street\altaffilmark{18},
Y. Tsapras\altaffilmark{19},
E. Bachelet\altaffilmark{18,20},
M. Dominik\altaffilmark{21,d},
D. M. Bramich\altaffilmark{20},
Keith~Horne\altaffilmark{21},
C. Snodgrass\altaffilmark{22},
I. A. Steele\altaffilmark{23},
J. Menzies\altaffilmark{24},
R. Figuera Jaimes\altaffilmark{21,25},
J.~Wambsganss\altaffilmark{19},
R. Schmidt\altaffilmark{19},
A. Cassan\altaffilmark{26},
C. Ranc\altaffilmark{26},
S. Mao\altaffilmark{27,28,29},
Subo Dong\altaffilmark{30},\\
(RoboNet)\\
and\\
G. D'Ago\altaffilmark{10},
G. Scarpetta\altaffilmark{5,10},
P. Verma\altaffilmark{10},
U.G. J{\o}rgensen\altaffilmark{8},
E. Kerins\altaffilmark{29},
J. Skottfelt\altaffilmark{8,31}\\
(MiNDSTEp) \\
}
\altaffiltext{1}{Jet Propulsion Laboratory, California Institute of Technology, 4800 Oak Grove Drive, Pasadena, CA 91109, USA}
\altaffiltext{2}{Warsaw University Observatory, Al.~Ujazdowskie~4, 00-478~Warszawa,Poland}
\altaffiltext{3}{Department of Astronomy, Ohio State University, 140 W. 18th Ave., Columbus, OH  43210, USA}
\altaffiltext{4}{Department of Physics, Chungbuk National University, Cheongju 361-763, Republic of Korea}
\altaffiltext{5}{Dipartimento di Fisica ``E. R. Caianiello'', Universit\`a di Salerno, Via Giovanni Paolo II, 84084 Fisciano (SA),\ Italy}
\altaffiltext{6}{Istituto Nazionale di Fisica Nucleare, Sezione di Napoli, Italy}
\altaffiltext{7}{School of Physics and Astronomy, Tel-Aviv University, Tel-Aviv 69978, Israel}
\altaffiltext{8}{Niels Bohr Institute \& Centre for Star and Planet Formation, University of Copenhagen, {\O}ster Voldgade 5, 1350 Copenhagen K, Denmark}
\altaffiltext{9}{NASA Exoplanet Science Institute, MS 100-22, California Institute of Technology, Pasadena, CA 91125, USA}
\altaffiltext{10}{Istituto Internazionale per gli Alti Studi Scientifici (IIASS), Via G. Pellegrino 19, 84019 Vietri sul Mare (SA), Italy}
\altaffiltext{11}{{\it Spitzer}, Science Center, MS 220-6, California Institute of Technology,Pasadena, CA, USA}
\altaffiltext{12}{UKIRT, 660 N. A‘ohoku Place, University Park, Hilo, HI 96720, USA}
\altaffiltext{13}{Harvard-Smithsonian Center for Astrophysics, 60 Garden St., Cambridge, MA 02138, USA}
\altaffiltext{14}{Department of Physics, University of Warwick, Gibbet Hill Road, Coventry, CV4 7AL, UK}
\altaffiltext{15}{University of Canterbury, Department of Physics and Astronomy, Private Bag 4800, Christchurch 8020, New Zealand}
\altaffiltext{16}{Korea Astronomy and Space Science Institute, Daejon 305-348, Republic of Korea}
\altaffiltext{17}{School of Space Research, Kyung Hee University, Yongin 446-701, Republic of Korea}
\altaffiltext{18}{Las Cumbres Observatory Global Telescope Network, 6740 Cortona Drive, suite 102, Goleta, CA 93117, USA}
\altaffiltext{19}{Astronomisches Rechen-Institut, Zentrum f{\"u}r Astronomie der Universit{\"a}t Heidelberg (ZAH), 69120 Heidelberg, Germany}
\altaffiltext{20}{Qatar Environment and Energy Research Institute, Qatar Foundation, P.O. Box 5825, Doha, Qatar}
\altaffiltext{21}{SUPA, School of Physics \& Astronomy, University of St Andrews, North Haugh, St Andrews KY16 9SS, UK}
\altaffiltext{22}{Planetary and Space Sciences, Department of Physical Sciences, The Open University, Milton Keynes, MK7 6AA, UK}
\altaffiltext{23}{Astrophysics Research Institute, Liverpool John Moores University, Liverpool CH41 1LD, UK}
\altaffiltext{24}{South African Astronomical Observatory, PO Box 9, Observatory 7935, South Africa}
\altaffiltext{25}{European Southern Observatory, Karl-Schwarzschild-Str. 2, 85748 Garching bei M\"unchen, Germany}
\altaffiltext{26}{Sorbonne Universit\'es, UPMC Univ Paris 6 et CNRS, UMR 7095, Institut d'Astrophysique de Paris, 98 bis bd Arago, 75014 Paris, France}
\altaffiltext{27}{Physics Department and Tsinghua Centre for Astrophysics, Tsinghua University, Beijing 100084, China}
\altaffiltext{28}{National Astronomical Observatories, Chinese Academy of Sciences, 20A Datun Road, Chaoyang District, Beijing 100012, China}
\altaffiltext{29}{Jodrell Bank Centre for Astrophysics, School of Physics and Astronomy, University of Manchester, Oxford Road, Manchester M13~9PL, UK}
\altaffiltext{30}{Kavli Institute for Astronomy and Astrophysics, Peking University, Yi He Yuan Road 5, Hai Dian District, Beijing 100871, China}
\altaffiltext{31}{Centre for Electronic Imaging, Department of Physical Sciences, The Open University, Milton Keynes, MK7 6AA, UK}
\altaffiltext{a}{NASA Postdoctoral Program Fellow}
\altaffiltext{b}{Sagan Visiting Fellow}
\altaffiltext{c}{Sagan Fellow}
\altaffiltext{d}{Royal Society University Research Fellow}

\begin{abstract}
We report the detection and mass measurement of a binary
lens OGLE-2015-BLG-1285La,b, 
with the more massive component having $M_1>1.35\,M_\odot$ (80\% probability).  
A main-sequence star in this mass range is ruled out by limits on
blue light, meaning that a primary in this mass range must be a neutron
star or black hole.
The system has a projected separation $r_\perp= 6.1\pm 0.4\,\au$ and
lies in the Galactic bulge.
These measurements are based on the ``microlens parallax''
effect, i.e., comparing the microlensing
light curve as seen from {\it Spitzer}, which lay at $1.25\,\au$
projected from Earth, to the light curves from four ground-based surveys,
three in the optical and one in the near infrared.  Future
adaptive optics imaging of the companion by 30m class telescopes
will yield a much more accurate measurement of the primary mass.
This discovery both opens the path and defines the challenges to
detecting and characterizing black holes and neutron stars in
wide binaries, with either dark or luminous companions.  In particular,
we discuss lessons that can be applied to future {\it Spitzer} and
{\it Kepler} K2 microlensing parallax observations.

\end{abstract}


\section{{Introduction}
\label{sec:intro}}

All known stellar-mass black holes (BHs) are in close binary systems.
Presumably this is a selection effect induced by the fact that these
BHs have been detected via X-ray emission generated by accretion from a close
companion.  In the near future, the Gaia satellite may detect (or
place interesting limits upon) BHs in wider binary systems with
main-sequence companions at 
semi-major axes $0.2\la a/\au\la 5$.  However, gravitational microlensing
appears to be the only way to systematically study the populations
of isolated BHs and BHs in well-separated binaries with dark (BH or
neutron star (NS)) companions. This is because, in the absence of a 
luminous companion,
BHs do not generate photometric signatures (except possibly
when accreting from the interstellar medium), and gravitational lensing
is unique in its ability to detect objects based solely on their
gravitational field \citep{einstein36}.

\citet{gould00a} estimated that almost 1\% of microlensing events
observed toward the Galactic bulge are due to BH lenses, with another
3\% due to NS lenses.  However,
even if this estimate is correct, not a single such lens has
yet been definitively identified.  The reason is quite straightforward:
the principal observable in microlensing events, the Einstein timescale
$t_\e$, is a combination of three physical properties of the lens-source
system,
\begin{equation}
t_\e = {\theta_\e\over\mu_\geo};
\qquad
\theta_\e^2\equiv \kappa M \pi_\rel;
\qquad
\kappa\equiv {4 G\over c^2\au}\simeq 8.14{{\rm mas}\over M_\odot}.
\label{eqn:tedef}
\end{equation}
Here $\theta_\e$ is the angular Einstein radius,
$\pi_\rel = \au(D_L^{-1}-D_S^{-1})$ is the lens-source relative parallax, and
$\mu_\geo$ is the lens-source relative proper motion in the Earth frame.
While the source distance $D_S$ is usually well-known, none of the
other variables in these equations are routinely measured.  Hence,
simple parameter counting implies that two other parameters must
be measured to determine $M$ and $D_L$.

Two quantities that can in principle be measured are the angular
Einstein radius $\theta_\e$ and the microlens parallax $\bpi_\e$,
whose magnitude is $\pi_\e = \pi_\rel/\theta_\e$ and whose direction
is that of the lens-source relative motion (see \citealt{gouldhorne}
Figure 1 for a didactic discussion).  If these two quantities
can be measured, then the physical parameters can be disentangled
\citep{gould92,gould00b}
\begin{equation}
M = {\theta_\e\over\kappa \pi_\e};
\qquad
\pi_\rel = \pi_\e\theta_\e;
\qquad
\bmu_\geo = {\theta_\e\over t_\e}{\bpi_{\e,\geo}\over\pi_\e}.
\label{eqn:meqn}
\end{equation}

For the great majority of microlensing events, neither $\pi_\e$ nor
$\theta_\e$ is measured.  However, for the special case of BH lenses,
the Einstein timescales 
$t_\e=\theta_\e/\mu\propto M^{1/2}$ tend to be long, and
this greatly enhances the prospects to measure $\pi_\e$ via Earth's
orbital motion
\citep{smith03,gould04}.  Hence, the main method that has
been recognized for identifying BHs and measuring their masses
using microlensing has been to attempt to measure $\theta_\e$ for these
long-timescale BH candidates.

However, this same large Einstein radius
$\theta_\e\propto M^{1/2}$ diminishes the already tiny chance 
$(p=\rho=\theta_*/\theta_\e)$ that the lens will transit a source of
radius $\theta_*$, which is the principal method by which $\theta_\e$
can be measured for isolated lenses \citep{gould94a,ob03262,gouldyee13}.

More than a decade ago, when microlensing event detections were almost
two orders of magnitude less frequent than today, three BH candidates
were identified, MACHO-98-6, MACHO-96-5, and MACHO-99-22.
\citet{poindexter05} subsequently showed that these had, respectively,
low, medium, and high probabilities to be BH lenses.  However, none
were either confirmed or rejected as BHs because there was no
measurement of $\theta_\e$.  The main idea to measure $\theta_\e$
for isolated BHs is astrometric microlensing, which takes advantage of
the fact that the displacement of the image centroid from the true
source positions is directly proportional to $\theta_\e$
\citep{walker95,hnp95,my95}.  Such
measurements are being pursued by several groups although they
are challenging with today's facilities.  However, they may become
routine in the future with {\it WFIRST} \citep{gouldyee14}.

Black holes in relatively wide ($\sim 1$--10 AU) binaries open a
second, less explored path toward BH microlensing mass measurements.
In contrast to isolated lenses, binaries often have large caustic
structures, which greatly increases the probability that the source
will transit these structures.  Because the magnification formally
diverges to infinity at a caustic, a caustic crossing permits a measurement of 
$\rho=\theta_*/\theta_\e$, the ratio of the angular source size to
the angular Einstein radius, since the observed magnification will be affected
by the finite size of the source.  Because $\theta_*$ is almost always easily
measured \citep{ob03262}, this yields $\theta_\e$.  Of course,
such measurements require that the events actually be monitored
during the typically brief (few hour) caustic crossings, which
in the past has required either good luck or very aggressive followup
observations.  This situation may change with the ramp up of
modern surveys like OGLE-IV and KMTNet, which monitor wide fields
at greater than 1/hour cadence for 13 and 16 square degrees, respectively.

To date, however, no such microlensing BH binaries have been discovered.
Part of the problem has certainly been missed caustic crossings, but
another part is that the entire ``paradigm'' outlined above actually
applies mainly to BHs in the Galactic disk.  From Equation~(\ref{eqn:meqn}),
the microlens parallax $\pi_\e=\sqrt{\pi_\rel/\kappa M}$.  Hence, $\pi_\e$
tends to be small for BHs.  If, in addition, the lens is in the Galactic
bulge, e.g., $\pi_\rel = 0.01\,\mas$, then an $M=5\,M_\odot$ BH would
have $\pi_\e=0.016$.  Such small parallaxes are difficult to detect,
and especially to reliably measure, from the ground.  Moreover, at a
proper motion $\mu=4\,\masyr$ typical of bulge lenses, the timescale
would be only $t_\e = 60\,$days, which is not exceptionally long.
Thus Galactic bulge BHs, including BH binaries
are much more difficult to detect and measure from the ground
than Galactic disk BHs.

Space-based microlensing parallax is well placed to meet these challenges.
Rather than relying on the fairly slowly accelerated motion of a single
observer on Earth \citep{gould92}, space-based microlensing directly
compares contemporaneous lightcurves from two well-separated observers
\citep{refsdal66,gould94b}.  For an Earth-satellite separation (projected
on the sky) $\bf D_\perp$, the microlensing parallax is approximately
given by
\begin{equation}
\bpi_\e = {\au\over D_\perp}(\Delta\tau,\Delta\beta);
\qquad \Delta\tau = {t_{0,\oplus} - t_{0,\rm sat}\over t_\e};
\qquad \Delta\beta = \pm u_{0,\oplus} - \pm u_{0,\rm sat},
\label{eqn:pieframe}
\end{equation}
where the subscripts indicate parameters as measured from Earth
and the satellite.  Here, $(t_0,u_0,t_\e)$ are the standard point-lens
microlensing parameters: time of maximum\footnote{For binary lenses, this
is replaced by time of closest approach to some fiducial point in the
lens geometry, which is usually not in fact the maximum.}, 
impact parameter, and timescale.
For binary microlensing, these form a subset of a larger parameterization,
but the formula remains valid.  For point-lens events, precisions at
the level $\sigma(\pi_\e)<0.01$ required for bulge-BH mass measurements
have been achieved in practice \citep{21event}.  For binaries, the
challenge is greater because of an intrinsic asymmetry in the
measurement of $(\Delta\tau,\Delta\beta)$ \citep{graff02}:  one
linear combination 
\begin{equation}
\Delta t_{\rm cc} = (\Delta\tau + \Delta\beta\cot\phi)t_\e,
\label{eqn:deltat}
\end{equation}
can be measured with exquisite precision from the difference in
caustic-crossing times $\Delta t_{\rm cc}$ seen from Earth and the satellite.
Here, $\phi$ is the angle between the source trajectory and the tangent
to the caustic.  However, the orthogonal combination is much more
difficult to measure.  We will discuss this challenge
in some detail in Sections~\ref{sec:parms+pie}, \ref{sec:future}, 
\ref{sec:break}, and \ref{sec:discuss}.

Nevertheless, the main difficulty is the availability of such 
parallax satellites,
which must be in solar orbit (or orbiting a solar system body that
is not itself orbiting Earth), capable of reasonably good photometry
in the crowded bulge fields, and, of course, allocated to microlensing
observations.

\citet{dong07} made the first such microlens parallax measurement
using the IRAC camera on {\it Spitzer} for the microlensing event
OGLE-2005-SMC-001. This was, in fact, a binary star lens and moreover 
the favored interpretation was a BH binary.  Unfortunately, however,
there was no caustic crossing, so this remains a candidate rather
than a confirmed detection.

In 2014, the Director allocated 100 hours of {\it Spitzer} time for
Galactic bulge observations with the specific aim of determining 
{\it Spitzer}'s viability as a microlensing parallax satellite.
Based on this successful test \citep{ob140124,ob140939,21event},
which included one mass measurement of a binary \citep{ob141050},
832 hours were awarded for 2015, which is a majority of the 38
days that bulge targets are visible from {\it Spitzer}
due to Sun-angle restrictions.

While the main focus of this program was to determine the Galactic
distribution of planets, and the main protocols for both
{\it Spitzer} and supporting ground-based observations were determined
on this basis \citep{yee15}, there was also a significant effort
to monitor binaries, exactly because of the possibility of mass measurements.

Here we report on the mass and distance measurements
of OGLE-2015-BLG-1285La,b.  The mass of the primary indicates that it is
most likely a NS or BH.  The system lies $<2^\circ$ in projection
from the Galactic center and is almost certainly a member of the
Galactic bulge population.
The field was specifically targeted by
OGLE for factor $\sim 5$ increased cadence to enable early alerts
that would permit timely {\it Spitzer} observations and to increase
the probability of resolving unexpected caustic crossings.  It was
further targeted as part of a $\sim 4\,\rm deg^2$ survey by
UKIRT and Wise observatories in Hawaii and Israel, respectively, in order
to both increase phase coverage and, in the former case, take advantage
of the capacity of IR observations to penetrate the relatively high
extinction in these fields.  The mass measurement of 
OGLE-2015-BLG-1285 is a specific product of these specially targeted
observations.

In Section~\ref{sec:obs} we discuss the observations, with emphasis
on {\it Spitzer} and the above mentioned special targeting.
In Section~\ref{sec:anal}, we present a microlens model
and demonstrate how the physical conclusions follow
from the light curve features and the source position on the color-magnitude
diagram (CMD).  In Section~\ref{sec:future}, we show that future proper-motion
measurement of the luminous component(s) of the binary lens will
yield a decisive mass measurement.  In Section~\ref{sec:break}, we discuss
the requirements for breaking similar degeneracies in future events using
microlensing data alone.
Finally, in Section~\ref{sec:discuss},
we discuss some other future prospects.

{\section{Observations}
\label{sec:obs}}

\subsection{Special Northern Bulge Fields}

The high extinction toward the Galactic plane in optical
surveys roughly splits the Galactic bulge into distinct northern and southern regions.
Microlensing surveys traditionally concentrate on southern bulge fields
close to the Galactic plane because the event rate there is high and the
extinction is relatively low.  This maximizes the number of detected
events with the high-quality light curves that are required for planet
detection.  It is of course understood that the northern bulge fields, being
roughly symmetric with the south, have just as many microlensing events.
However, prior to the 2015 {\it Spitzer} campaign, 
only one northern bulge field was targeted
for high-cadence observations: OGLE-IV BLG611, centered at 
$(\ell,b)=(0.33,2.82)$.
Although this field has a long heritage going back almost two decades to OGLE-II, 
no systematic study had ever been made as to which northern bulge
fields were the most profitable to target.

In the course of analyzing the 2014 {\it Spitzer} campaign, we realized
that {\it Spitzer} target selection was being heavily influenced by
optically-based microlensing alerts, whose distribution on the
sky was strongly impacted both directly and indirectly by 
the pattern of dust extinction.  That is, first, to the extent that high-extinction
fields are surveyed, it is more difficult to find and monitor events
because they are systematically fainter in the optical. Second, because
of this very fact, these fields tend to be monitored at lower cadence or
not at all.
By contrast,
{\it for any event that can be detected in the optical} (i.e., $A_I\la 4$)
{\it Spitzer} is essentially unaffected by the dust.  In 2015, therefore,
special efforts were made to counter this bias, which included
taking specific account of the extinction at each event location
\citep{yee15}.  In addition to these general measures, we also
identified several northern bulge fields for special observations,
including an $H$-band survey using UKIRT (Hawaii) and
an $I$-band survey from Wise (Israel).  One of the four OGLE
fields containing these regions was BLG611, which was already
being observed with hourly cadence.  But the other three fields
BLG653, BLG654, and BLG675 (cf., Fig.~15 in \citealt{ogleiv})
were raised from cadences of roughly
0.5/day to 2--3/day.

The UKIRT/Wise fields were selected using the procedures developed by
\citet{poleski15}, who showed that the product of the surface densities
of $I<20$ stars and clump stars is a good predictor of the microlensing
event rate.

\subsection{OGLE Alert and Observations}

On 2015 June 7 UT 19:39, the Optical Gravitational Lens Experiment (OGLE)
alerted the community to a new microlensing event OGLE-2015-BLG-1285
based on observations with the 1.4 deg$^2$ camera on its 
1.3m Warsaw Telescope at the Las Campanas Observatory in Chile \citep{ogleiv}
using its Early Warning System (EWS) real-time event detection
software \citep{ews1,ews2}.  Most observations were in $I$ band, but with
some $V$ band observations that are, in general, taken for source
characterization.  These are not used in the modeling.
At equatorial coordinates (17:39:23.75, $-27$:49:13.0) and
Galactic coordinates $(0.23,-1.75)$, this
event lies in OGLE field BLG675, with a nominal observing cadence of
roughly 2--3 times per night.

\subsection{{\it Spitzer} Observations}

OGLE-2015-BLG-1285 originally appeared to be a point-lens event.
The protocols and strategies for observing such events with 
{\it Spitzer} are reviewed in \citet{ob150966} and are discussed in
greater detail by \citet{yee15}.  In brief, targets
were submitted on Monday for observations on the following Thursday through Wednesday
for each of the six weeks of the {\it Spitzer} campaign.

OGLE-2015-BLG-1285 was selected for {\it Spitzer} observations on
Monday June 22 UT 13:33, i.e., for the fourth week of observations,
and at standard hourly cadence.  Given the time of the OGLE alert,
it could not have been selected for the first two weeks.  The {\it Spitzer}
team specifically considered this event during preparations for the
third week but found its predicted behavior too ambiguous to select it.
Even in the fourth week, it was considered highly risky but was chosen
specifically because it lay in a field covered by UKIRT and Wise, and
so would automatically receive good light curve
coverage.  Also, it was noted that
the source was probably a red giant and so about 100 times brighter
at $3.6\,\mu$m than typical targets, even though it was relatively
faint in the optical due to high extinction.

In the sixth (final) week, the cadence was increased to 4/day on the grounds
that it was apparently anomalous.  According to the protocols of \citet{yee15},
such increased cadence can be used to characterize the anomaly or increase
the precision of the parallax measurement, provided that the anomaly
and parallax are detectable without them.  Altogether, {\it Spitzer}
observed OGLE-2015-BLG-1285 38 times, with each epoch composed of six
30-second dithered exposures.

\subsection{Other Survey Observations}

The sky position of OGLE-2015-BLG-1285 was covered by three surveys
in addition to OGLE, namely UKIRT, Wise, and KMTNet.  As with OGLE,
the UKIRT and KMTNet observations were carried out without consideration of 
any known microlensing events in the field.  The Wise observation procedures
are discussed explicitly, below.

UKIRT observations were carried out with the wide-field NIR camera WFCAM, at
the UKIRT telescope on Mauna Kea, Hawaii. WFCAM uses four Rockwell Hawaii-II
HgCdTe detectors. The field
of view of each detector is $13.6'\times13.6'$ and the four arrays are
separated by gaps whose areas are 94\% of one detector.
The observations were in $H$ band, with each epoch composed of 
sixteen 5-second co-added dithered exposures (2 co-adds, 
2 jitter points and $2\times2$ microsteps).
The 18 survey fields were observed 5 times per night.  

The Wise group used the recently installed
Jay Baum Rich 0.71m Telescope (C28) at Wise Observatory in Israel,
equipped with a 1 $\rm deg^2$ camera.
The 4 survey fields were observed 5 times per night.
At the time, the C28 had some pointing problems,
and OGLE-2015-BLG-1285 was close to the edge of the survey field.
Hence, at the initiative of MF,
the Wise group decided to also monitor the event with the Wise 1m telescope
equipped with the PI camera.  Because these observations were triggered
solely to ensure coverage of an event that was in the survey field, 
with a cadence similar to that of the survey, we treat
these as ``survey'' observations even though they were taken with
a different telescope.  In fact, only the PI observations usefully
constrain the model, so we do not include the C28 data.

KMTNet is a new survey that employs $4\,\rm deg^2$ cameras at three sites:
CTIO/Chile, SAAO/South Africa, and SSO/Australia \citep{kmtnet}.
While the primary
goal of this survey is near-continuous observation of $16\,\rm deg^2$
in the southern bulge, it supported the {\it Spitzer} campaign by
obtaining data on another $40\,\rm deg^2$, with cadence of 1--2/day
at each telescope.  These lower-cadence fields included the location
of OGLE-2015-BLG-1285.

\subsection{Followup Observations}

Sustained followup observations were carried out by the RoboNet team using
5 telescopes
from the Las Cumbres Observatory Global Telescope (LCOGT) in Chile, South
Africa, and Australia, and by the Microlensing Follow Up Network ($\mu$FUN)
1.3m SMARTS telescope at CTIO.  By chance, these observations did not
cover the crucial bump in the light curve, while their coverage of
the wings adds only modestly to the survey coverage.  Hence, they
do not significantly influence the fits.  They are nevertheless
included for completeness.

The {\it Spitzer} team issued an anomaly alert for OGLE-2015-BLG-1285
on July 6 UT 14:43 (JD 7210.11)
based on a single OGLE data point that was posted on its web page.
The Salerno University 0.35m telescope responded to this alert 
when the event rose over Italy just 5 hours later.  These data
begin at the tail end of the caustic crossing.  They qualitatively
confirm the exit feature traced by the Wise 1m.  However, we do
not include them in the fit because the target brightness was
at the margin of obtaining reliable data.

Thus, the results reported here depend overwhelmingly on survey data.

\subsection{Data Reduction}

All ground-based data were reduced using standard algorithms.
Most data entering the main analysis
used variants of image subtraction \citep{alard98}.
CTIO-SMARTS, UKIRT, and Wise data were reduced using DoPhot \citep{dophot},
while the LCOGT data were processed using DanDIA \citet{Bramich2008}.

{\it Spitzer} data were reduced using a new algorithm 
(Calchi Novati 2015, in preparation) whose necessity
is discussed in \citet{yee15}.

{\section{Light Curve Analysis}
\label{sec:anal}}

{\subsection{Ground-based Light Curve}
\label{sec:gblc}}

The light curve contains only a single pronounced feature, which occurs
near the peak of a roughly symmetric and otherwise low-amplitude single-lens event, 
i.e., less than one magnitude above baseline just before and after
the sudden anomaly. See Figure~\ref{fig:lc}.
In order to properly estimate the masses and
projected separation of the components, as well as the distance to
the system, we must unambiguously characterize the binary geometry
from this single feature, combined with the more subtle variations
of the rest of the light curve.  From the fact that the {\it Spitzer}
and ground-based lightcurves are offset by only
\begin{equation}
\Delta t_{\rm peak} \equiv t_{\rm peak,\oplus}  - t_{\rm peak,sat} \sim 0.3\,\rm day,
\label{eqn:peak}
\end{equation}
(compared to the several week duration of the event)
we already know that the microlens parallax effects can be ignored
to first order for the ground-based light curve.  However,
this still leaves seven geometric
parameters to be determined $(t_0,u_0,t_\e,t_*,\alpha,s,q)$. 
Here, $(t_0,u_0,t_\e)$ are the three parameters of the underlying
single lens event, $(\alpha,s,q)$ are the three binary-lens parameters,
and $t_*\equiv \rho\,t_\e$ is the source crossing time, where 
$\rho\equiv \theta_*/\theta_\e$ is ratio of the angular source
size to the angular Einstein radius.  The three underlying point
lens parameters are, respectively,
the time of closest approach to some
fiducial point in the geometry, the impact parameter (normalized
to $\theta_\e$), and the Einstein timescale.
The three binary parameters are, respectively,
the angle of the source trajectory relative to the binary axis, the
projected binary separation (normalized to $\theta_\e$), and the binary
mass ratio.

How can we make an exhaustive search of such a large parameter space?
We begin by noting that the rough symmetry, dramatic outburst at peak,
and lack of significant dip within this peak, together imply that
the source is moving nearly perpendicular to the binary axis, and
that it intercepts a cusp (or two very close and 
roughly parallel caustics) on this axis.  
We initially ignore the second, rather special geometry.  Then, for 
each pair $(s,q)$ there
are either two cusps (for close and resonant binaries) or four cusps
(for wide binaries) on the binary axis.  Some cusps can be excluded
because perpendicular trajectories would yield pronounced
bumps as the source crossed or passed nearby to neighboring cusps.
For each of the remaining cusps for a given $(s,q)$, we first set the
origin of the coordinate system at the cusp (rather than the center of
mass or center of magnification, as is customary). With this parameterization,
$(t_0,u_0,t_*)$ are approximately uncorrelated so that at fixed $(s,q)$,
there remain only two parameters with significant correlations, $t_\e$ and $\alpha$.

We search for solutions using the Monte Carlo Markov chain technique.
For each set of trial parameters we employ contour integration \citep{gg97}
for points that either straddle or are very close to a caustic, using
10 annuli to allow for limb darkening.  For points that are further
from the caustics we progressively use the hexadecapole, quadrupole,
and monopole approximations \citep{pejcha09,gould08}.  We use linear
limb-darkening coefficients of $u_I=0.61$, $u_H=0.42$, and $u_{[3.6]}=0.28$ from 
\citet{claret00},
based on the source type derived from Figure~\ref{fig:cmd}.  In the last
case, we must extrapolate.  For each model geometry and each observatory, $i$,
we fit for a source flux $f_{s,i}$ and a blend flux $f_{b,i}$ that minimizes
the $\chi^2$ of the observed fluxes $F_{i,\rm obs}(t)$ relative to the
predicted fluxes
\begin{equation}
F_{i,\rm pre}(t) = f_{s,i}A(t;t_0,u_0,t_\e,t_*,\alpha,s,q) + f_{b,i} .
\label{eqn:fluxpre}
\end{equation}

We choose an
initial seed trajectory by 
$(t_0,u_0,t_\e,t_*,\alpha)= (7209.75,0,t_\e,0.45\,{\rm day},90^\circ)$,
where $t_\e=30\,$day or $t_\e=50\,$day as discussed immediately below.
For the first 100 trials, we allow only $u_0$ to vary because the
crossing may be 0.02--0.1 Einstein radii from the cusp
(i.e., $u_0=0$), depending
on the topology.  Then, because $(t_0,u_0,t_*)$ are approximately correct and
because there
are only two correlated parameters $(t_\e,\alpha)$, the Markov chain arrives
near the $\chi^2$ minimum very fast.

There are only two topologies that yield a competitive $\chi^2$.
This agrees with the results of two completely independent, and
generalized search algorithms (VB\footnote{http://www.fisica.unisa.it/gravitationAstrophysics/RTModel/2015/RTModel.htm}
and CH\footnote{http://astroph.chungbuk.ac.kr/~kmtnet/2015.html})
that do not make use of the
detailed features of the light curve outlined above.
In both topologies, the trajectory passes through a cusp of the caustic structure
associated with the lower-mass component. In the first, it passes
by the inner cusp of the lower-mass component, i.e., the cusp that lies
closer to the higher-mass component.  In the second, it passes by
either the outer cusp of this caustic, or essentially the same cusp
of a resonant binary, i.e., the cusp associated with the lower-mass component.
The first topology (see Figure~\ref{fig:caust} for several examples)
typically has timescales $t_\e\sim 30\,$days,
so we use this value in our seeds for this topology.  The second topology
has $t_\e\sim 50\,$days, so we seed with this value in those cases.
(While the second topology is not shown in Figure~\ref{fig:caust},
it would consist of the source passing roughly perpendicular to the
binary axis and right through the bottom-most cusp.)

The ``inner cusp'' topology is favored by $\Delta\chi^2=55$,
which is strong evidence in its support.  Figure~\ref{fig:inner} shows
$\chi^2(s,q)$ for this topology, and Figure~\ref{fig:caust} illustrates
a range of caustic morphologies drawn from this minimum.
In addition to having different topologies, the two solutions are characterized
by very different amounts of blend flux.  As we will see in the next section, this 
implies that the inner topology is strongly favored by another, 
independent argument based on astrometric and chromatic constraints.

{\subsection{Astrometric and Chromatic Constraints
\label{sec:astrochrom}}

Neither the position nor the color of the apparent source changes
perceptibly
as it increases its brightness by a factor $\sim 15$ during the event.
This is exactly what what one would expect for an unblended source
(inner topology) but requires extreme fine tuning for the outer
topology\footnote{We do not discuss this solution in detail because
we consider it ruled out.  However, for completeness we note that 
it has $(t_\e,t_*,\pi_{\e,N},\pi_{\e,E},\alpha,s,q)=
(46.3\,{\rm day},0.42\,{\rm day},0.000,0.004,90.1^\circ,1.56,7.2)$
and therefore $(M_1,M_2) = (13.7,1.9)\,M_\odot$.}
 ($f_b\simeq 0.68\,f_s$).

\subsubsection{Chromatic Constraint}

Figure~\ref{fig:cmd} is an $(I-H,I)$ CMD constructed by aligning 
OGLE $I$-band and UKIRT $H$-band data.  The centroid of the clump 
and the ``baseline object'' at the location of the source are marked.
We note that the $H$-band zero point is not fully calibrated, but
such calibration is not needed in the present context because all
results are derived from relative photometry.  

Model-independent regression of $H$-band
on $I$-band flux during the event yields 
\begin{equation}
\delta(I-H)\equiv (I-H)_s - (I-H)_b \simeq (-0.009\pm 0.008)\times 
10^{0.4(I_b-I_{\rm base})} .
\label{eqn:ihregress}
\end{equation}

Regardless of the degree of blending,
the source color is the same as that of the baseline object shown
in Figure~\ref{fig:cmd}, whose position relative to the clump is
\begin{equation}
\Delta [(I-H),I] = [(I-H),I]_{\rm base} - [(I-H),I]_{\rm clump} = (0.02,0.10).
\label{eqn:ih}
\end{equation}
Since $(I-H)_s-(I-H)_{\rm base}\simeq 0$ to high precision, this permits us to
derive
\begin{equation}
\theta_* = 6.01\,\muas\,\sqrt{f_s\over f_{\rm base}},
\label{eqn:thetastar}
\end{equation}
using the standard method of offset from the clump \citep{ob03262}.
That is, for a star at the center of the bulge clump 
$((V-I),I)_{s,0}=(1.06,14.46)$ \citep{bensby13,nataf13}, the source
size (if unblended) would be $\theta_*=6.17\,\muas$ \citep{kervella04,bb88}.  
Since the baseline
object is fainter by 0.1 mag, this number is reduced by $10^{-0.2\times 0.1}$.
And since it is redder by $\Delta(I-H)=0.02$, its surface brightness
is lower by 4\% \citep{boyajian14}, so 2\% bigger at fixed magnitude.
Hence, $\theta_*=6.01\,\muas$, which can then be scaled to the source
flux as in Equation~(\ref{eqn:thetastar}).

\subsubsection{Astrometric Constraint}

The source position is measured from the difference image of the
event at peak magnification relative to baseline.  Since there
are no stars except the source in such a difference image, the
offset can be measured with great precision, $\sigma= 0.02$ OGLE
pixels (each 260 mas).  The baseline position is measured from a
stack of excellent images and has a precision $\sigma= 0.045\,$pixels.
Hence, the combined uncertainty in the difference between the positions
of the source and the baseline object is
 $\sigma=0.05\,$pixels or 13 mas.  The actual
difference in positions is 0.05 pixels in each direction.  This
is consistent at the $1\,\sigma$ level 
with the hypothesis of an unblended source.

\subsubsection{Application to Outer Cusp Topology Solutions}

However, for the outer cusp solution  ($f_b\simeq 0.68\,f_s$), these constraints
together imply that by
chance another red giant of very similar color to the source
and less than 1 mag below the
clump lies within a few tens of mas of the source.
There are only four ways that this can happen.  Either
the blend is directly associated with the source (i.e., they form
a red-giant binary), directly associated with the lens (red giant companion
to the binary lens), a component of the lens,
or the additional red giant is an unassociated star that lies projected
within 40 mas of the source.

The prior probability for the first option (red-giant binary source)
can be evaluated in two steps.  First, the fraction of G dwarfs
(the progenitors of bulge clump stars) with companions within 
$0.9<M_{\rm comp}/M_{\rm prim}<1.0$ is about 3\% \citep{dm91}.
On the other hand, for a 10 Gyr solar-mass solar-metallicity
isochrone, the mass difference between 25 and 250 solar luminosities
(encompassing a conservatively large range of the giant branch) is
$M(250\,L_\odot)-M(25\,L_\odot)= 1.8\times 10^{-3}M_\odot$. Hence, the
prior probability is $0.03\times (1.8\times 10^{-3}/0.1)\sim 5\times 10^{-4}$.

The prior probability of the second option (red giant companion to the
binary lens) is somewhat smaller than this, 
since the conditional probability for
a tertiary given a few-AU binary is smaller than the probability of
a companion given the presence of a single star.

The prior probability of the third option depends on the details
of the solution.  However, it is a maximum if the companion mass
is of order one solar mass.  In this case, the above calculation can
be applied but without the binarity factor, i.e., 1.8\%.  Again, this
assumes that the solution predicts $M_1\sim 1\,M_\odot$ or $M_2\sim 1\,M_\odot$.  
Otherwise, the probability is close to zero.

The probability of chance projection is even smaller.  First, if the
source is separated from the centroid by $<40\,\mas$ ($2\,\sigma$ limit), 
then it is separated
from the blend by $<100\,\mas$ (assuming roughly equal brightness).
In this field, the density of stars that are no more than 1 mag below
the clump and within $|\Delta(I-H)|<0.05$ of a given color (in this
case, the source color) is $16\,\rm arcmin^{-2}$.  
Therefore the probability of such a projection is
$\pi(100\,\mas)^2\times 16\,\rm arcmin^{-2} \sim 10^{-4}$.  In sum, the
total probability is about 2\% for the case that one of the components
is about $1\,M_\odot$ and $<0.1\%$ otherwise.

{\subsubsection{Application to Putative Blue Lenses}
\label{sec:blue}}

One may also apply the color constraint to the inner-cusp solutions,
for which the blending is constrained by the fit to be small, but
may not be exactly zero.  In particular, having one or both components 
be main-sequence stars would in itself be consistent with upper
limits on the blended flux.  However, the color constraint implies that
any such stars must be quite faint. For example, consider 
F type stars, e.g., $(I-H)_0\sim 0.4$.  Then $\Delta(I-H)\sim 0.7$,
which implies $I_b-I_s\ga 3.5$ at $3\,\sigma$ confidence.  This
essentially rules out $M>1.35\,M_\odot$ main-sequence stars.

{\subsection{Ground-Only Microlens Parameters
\label{sec:parms}}

Figure~4 shows the $\chi^2(s,q)$ surface for the ``inner topology''.
The black curve shows the boundary between wide-binary and resonant caustic
topologies \citep{erdl98},
\begin{equation}
s^2 = {(1+q^{1/3})^3\over 1+q}.
\label{eqn:bound}
\end{equation}
That is, while the minimum does lie in the wide-binary caustic topology,
the $1\,\sigma$ contour crosses the boundary into resonant caustics.
Of course, they cannot cross very far because then the ``neck'' connecting
the two formerly separate wide-binary caustics would widen, 
leading to a dip in the middle of the bump, which is not seen in 
Figure~\ref{fig:lc}.  Hence, it has proved unnecessary to make an independent
search of this narrow-neck resonant topology, mentioned at the
beginning of Section~\ref{sec:anal},
since it is contiguous with the inner cusp wide-binary topology.

Table \ref{tab:model} shows the best-fit microlens parameters 
(including $\bpi_\e$, which is discussed below) and their 68\%
confidence intervals, derived from the MCMC chain density,
leading to $\theta_\e\sim0.42\,\mas$.
The $\chi^2$ surface is relatively far from parabolic,
and there are non-linear correlations among the parameters,
thus these values should not be considered as standard errors.

{\subsection{Microlens Parallax From {\it Spitzer} Light Curve}
\label{sec:parms+pie}}

The microlens parallax $\bpi_\e$ of OGLE-2015-BLG-1285 is quite
small and can only be measured because of the long baseline 
$D_\perp\sim 1.25\,\au$ provided by {\it Spitzer}.  It is straightforward
to incorporate {\it Spitzer} data into the binary lens fit (e.g., 
\citealt{ob141050}), add in two parameters for $\bpi_\e$, and report the
result.  However, it is also important to gain a physical understanding of
how the features of the {\it Spitzer} light curve act to constrain the parallax.

The strongest constraining feature is the time of peak, which is
$\Delta t_{\rm peak} = 0.3\,$days earlier than the ground-based peak
(Equation~(\ref{eqn:peak})).  Since the {\it Spitzer} and ground peaks are
both due to the source crossing the binary axis, we find from simple
geometry that
\begin{equation}
{\Delta t_{\rm peak}\over t_\e} = \Delta \tau +\Delta\beta\cot\alpha .
\label{eqn:pieconstraint}
\end{equation}
Note that this is identical to the generic Equation~(\ref{eqn:deltat}),
but with $\phi\rightarrow\alpha$, which follows from the fact that
the caustic is tangent/parallel to the binary axis.
Equation~(\ref{eqn:pieconstraint}) 
can be combined with Equation~(\ref{eqn:pieframe}) to yield
\begin{equation}
\pi_{\e,\Delta\tau} + \cot\alpha\pi_{\e,\Delta\beta} = {\au\over D_\perp}\,
{\Delta t_{\rm peak}\over t_\e} \simeq 0.008,
\label{eqn:pieconstraint2}
\end{equation}
where the subscripts $\Delta\tau$ and $\Delta\beta$ refer to the
direction parallel and perpendicular to the projected separation $\bf D_\perp$,
which are close to east and north, respectively.

Then, for any given geometry that is consistent with ground-based
data, the position of $\bpi_\e$ within the one-dimensional (1-D) space
defined by Equation~(\ref{eqn:pieconstraint2}) must be determined primarily
by the data points on the approach to the cusp.  That is, a large value of
$\Delta\beta$ would imply that the source crosses the binary axis substantially
closer in (or farther out) as seen from {\it Spitzer} than Earth,
leading to a somewhat different cusp-approach morphology.  However, because 
these differences are not expected to be large, and because {\it Spitzer}
observed with only daily cadence, we expect these constraints to
be much weaker than those in Equation~(\ref{eqn:pieconstraint}).
Hence we expect elongated error contours in the $\bpi_\e$ plane.

Figure~\ref{fig:pie} shows these contours.  For binary lenses with
ground-based data there is often a degeneracy that takes
$(u_0,\alpha,\pi_{\e,N})\rightarrow -(u_0,\alpha,\pi_{\e,N})$
\citep{ob09020}, which is sometimes dubbed the ``ecliptic degeneracy''
because it is exact on the ecliptic.  The north component of $\bpi_\e$ is
singled out because the ecliptic happens to run east-west for
bulge fields.  One expects this degeneracy to be preserved for
{\it Spitzer} because it also lies very close to the ecliptic.
Figure~\ref{fig:pie} shows that this is indeed the case.

Because $M=\theta_\e/\kappa\pi_\e$, and $\theta_\e$ varies very
little between viable solutions, the total mass $M$ tracks 
(inversely) the $\pi_\e$ values in Figure~\ref{fig:pie} very
well.  However, because a range of mass ratios $q$ are permitted
(see Figure~\ref{fig:inner}), the probability contours for the two
component masses, $M_1$ and $M_2$, (Figure~\ref{fig:mass})
are less 1-D than the $\bpi_\e$ contours.
These contours were calculated using the MCMC chain density,
and accounting for the Jacobian of the transformation between the MCMC variables
and the physical quantities
as derived in \citet{batista11} (see their Eq. 17-18).
This transformation requires priors on the mass function and the local density of lenses
(see definition in \citealt{batista11}).
We assume that the mass function is uniform in $\log(M)$, 
and evaluate from the CMD the bulge distance-modulus dispersion towards the event, $\sigma_{\rm DM} = 0.26$
\citep{nataf13}.

Finally, we note that from the definition of $\theta_\e$ 
(Equation~(\ref{eqn:tedef})), and the fact that $\theta_\e$ is essentially
the same in all solutions, we have
\begin{equation}
\pi_\rel = 10\,\muas
\biggl({\theta_\e\over 0.41\,\mas}\biggr)^2\biggl({M\over 2\,M_\odot}
\biggr)^{-1}
\label{eqn:pirel}
\end{equation}
Since the source is nearly at the center of the clump, it is
almost certainly in the bulge, with distance $D_S\sim 8\,\kpc$.
Equation~(\ref{eqn:pirel}) then implies 
$D_S-D_L\simeq 0.6\,\kpc(M/2\,M_\odot)^{-1}$, i.e., that the lens is also in
the bulge.

Taking account of the $\sim 8\%$ error in $\theta_*$ (and so $\theta_\e$),
there is an additional $1\,\sigma$ error in $M$ and $\pi_\rel$ of 8\% that
must be added in quadrature.  However, this is smaller
than the errors that propagate directly from fitting the light curve.

Figure~\ref{fig:diffm} shows the probability distribution of the primary
mass $M_1$, which is peaked at $M_1\sim 2.0\,M_\odot$ (black). The fraction of
the curve area $M>1.35\,M_\odot$ (typical NS mass) is 80\%.  As mentioned
in Section~\ref{sec:blue}, main-sequence stars in this mass range
are ruled out by the chromatic constraints.  Together, these imply that
this primary is a massive-remnant (NS or BH) candidate.

Table \ref{tab:physical} summarises the median and the 68\%
confidence intervals for the physical parameters of the binary system.
As for Table \ref{tab:model}, we warn about the non-gaussianity and non-linear correlations
among the MCMC variables, which  are also reflected in the uncertainties 
on the physical parameters.

{\section{Future Mass Determination from Proper Motion 
Measurement}
\label{sec:future}}

The nature of OGLE-2015-BLG-1285La, i.e., whether it is a massive
remnant and if so whether it is a BH or NS, can be decisively resolved
by a proper-motion measurement of its companion, OGLE-2015-BLG-1285Lb,
whose mass implies that it is almost certainly a luminous low-mass main-sequence star
or a white dwarf (see Figure~\ref{fig:mass}).  This is illustrated
in Figure~\ref{fig:pie}, where we show a hypothetical future
measurement of the direction of lens-source relative proper motion
$\bmu$ with an error of either $3^\circ$ of $1^\circ$.  Because $\pi_\rel$ is very 
small, $\bmu=\Delta\btheta/\Delta t$, where $\Delta\btheta$ is the
observed lens-source separation at a future epoch $t_{\rm peak}+\Delta t$.

In order to make such a measurement, it is of course necessary for
the source and lens to separate.  Since their proper motion is known
$\mu=\theta_\e/t_\e\simeq 4.8\,\masyr$, the wait time depends primarily
on the resolution of the telescope.  \citet{batista15} were able to
resolve the source and lens of OGLE-2005-BLG-169 using Keck when they were 
separated by only 61 mas.  However, in that case the source and lens
had comparable brightness, whereas here they have a flux ratio
$f_l/f_s\la 1\%$.  Therefore, it is likely that 15 years would be
required with present instruments.  However, during this interval,
it is likely that 30m class telescopes with AO capability will
come on line. By the time that they do, this measurement will
already be quite feasible.  For example, for separations of 
$\Delta\theta\sim 50\,\mas$ and limiting resolution FWHM $\sim 11\,\mas$
(e.g., in $J$-band on the Giant Magellan Telescope),  a $1^\circ$ measurement
would require that the companion be centroided to 8\% of the FWHM
in the transverse direction.  Improvements to higher precision would
be considerably more difficult because the orbital motion of the secondary
about the center of mass is of order $0.5\,\mas$.

Figure~\ref{fig:diffm} shows the impact of such a proper motion measurement
on the estimate of the primary mass for $\pm 3^\circ$ (brown)  and
$\pm 1^\circ$ (magenta) errors, respectively.  Such a measurement would automatically
detect (or rule out) light from a main-sequence (or giant-branch) primary.
If the primary is indeed dark, then
the detection of light from the secondary would yield a mass estimate of 
that body, which would in principle constrain the solution and so further
constrain the mass of the primary. However, in practice we find that this
does not lead to significant further improvement beyond what can be achieved
with the proper motion measurement alone.

Finally, we note that \citet{gould14} has investigated the
problem of turning 1-D parallax measurements into 2-D parallaxes
via such proper motion measurements and shown that, in general,
there is a discrete degeneracy induced by the difference between the
geocentric frame of the $\bpi_\e$ measurement and the
heliocentric frame of the $\bmu$ measurement.  While the physical
origins of the 1-D degeneracy are completely different in the present case,
the mathematics, leading to a quadratic equation, are identical.
However, in the language
of that paper, $A\ll 1$ and $C\ll 1$, so the ``alternate'' solution
(Equation~(10) of \citealt{gould14}) is so large as to be easily
ruled out by the light curve.  Hence, there is no degeneracy.

{\section{The General Problem of
Breaking the 1-D Binary-Lens Parallax Degeneracy}
\label{sec:break}}

As discussed in Section~\ref{sec:intro}, events that have only a single
caustic crossing that is monitored from both Earth and a satellite
generically suffer from a 1-D degeneracy in $\bpi_\e$.  See 
Equations~(\ref{eqn:deltat}) and (\ref{eqn:pieconstraint}).
Breaking this degeneracy, in particular breaking it at the high
precision required for bulge-BH mass measurements (with their very
small $\pi_\e$), requires additional high-precision information.
In Section~(\ref{sec:future}) we showed that for the specific
case of OGLE-2015-BLG-1285, it will be possible to break this 
degeneracy with future proper motion measurements, although
this still will not deliver a precise mass measurement. Moreover,
first, this requires waiting many years, and second it will only
be possible because the secondary (or possibly primary) is luminous.
Hence, this method will not be applicable to binaries composed of two
remnants, which is the unique province of microlensing (and arguably
the most interesting case). 

Thus, it is of interest to understand how this degeneracy can be
broken from microlensing data alone, and in particular, what
can be done to modify current experimental protocols to increase the
chances of success.
Generically, this requires an additional feature in the light curve
that is monitored from both Earth and the satellite and whose appearance
from the satellite is predicted to be a function of the Earth-satellite
separation vector projected onto the Einstein ring.  For example,
while OGLE-2014-BLG-1050 does suffer from a 1-D degeneracy
(Figure 3a of \citealt{ob141050}), it is much less severe than that of 
OGLE-2015-BLG-1285.  This is primarily because the source trajectory
skirts the interior edge of the caustic (Figure 2a of \citealt{ob141050}),
which has 
significant structure.  Another widespread feature of binary light curves
that could provide such leverage is a cusp-approach ``bump'', which
often occurs shortly after a caustic exit.

However, by far the best structure would be a second caustic crossing.
In general, if there is a caustic entrance, there must be a caustic exit.
This was not actually true of OGLE-2015-BLG-1285 because the source
was much bigger than the separation between the caustics, so the entrance
and exit combined to form a single bump.   
However, if the source had been a main-sequence
star, i.e., 10 times smaller, there would have been a dip between the
the entrance and exit, whose duration would probe $\bpi_\e$ in the
orthogonal direction (although not with our data set, which has only
two {\it Spitzer} points over the entire cusp crossing).

More generally however, if there are two independent crossings with
measured $\Delta t_{1,2}$ and well-determined crossing angles $\phi_{1,2}$
(derived from the model), then one easily finds
\begin{equation}
\sigma(\Delta\tau) = {\sqrt{\cot^2\phi_1 + \cot^2\phi_2}
\over|\cot\phi_1-\cot\phi_2|}\,{\sigma(\Delta t)\over t_\e};
\qquad
\sigma(\Delta\beta) = {\sqrt{2}
\over|\cot\phi_1-\cot\phi_2|}\,{\sigma(\Delta t)\over t_\e},
\label{eqn:twocaust}
\end{equation}
where we have assumed that the errors $\sigma(\Delta t_{1,2})$
in the determinations of $\Delta t_{1,2}$ are the same.  The most
challenging case is clearly $|\Delta\phi|\equiv |\phi_1-\phi_2|\ll 1$,
for which
\begin{equation}
\sigma(\Delta\tau) \rightarrow {\sin(2\phi)\over\sqrt{2}|\Delta\phi|}
\,{\sigma(\Delta t)\over t_\e};
\qquad
\sigma(\Delta\beta)\rightarrow {1-\cos(2\phi)\over\sqrt{2}|\Delta\phi|}
\,{\sigma(\Delta t)\over t_\e}.
\label{eqn:twocaust2}
\end{equation}
It is
quite plausible to reach errors of $\sim 0.01$ days for these
time offsets, so that even assuming that the first factors in
Equations~(\ref{eqn:twocaust}) and (\ref{eqn:twocaust2}) are
of order $\sim 10$, the errors $\sigma(\Delta\tau)$ and
$\sigma(\Delta\beta)$ would be only $\sim 0.002$ 
for a $t_\e\sim 50\,$day event.  Thus, coverage of two independent
caustic crossings is by far the best method to measure $\bpi_\e$
for binaries.  We discuss how this can be achieved in practice
in the next section.

{\section{Discussion}
\label{sec:discuss}}

While it is not yet known whether OGLE-2015-BLG-1285La,b contains a massive
remnant, its example at least shows that such detections are possible.
The main challenge to detecting more of these systems is simply to
monitor a large number of targets, including both light curves that
already have clear binary signatures, but also those (like OGLE-2015-BLG-1285)
that erupt with these signatures unexpectedly.  {\it Spitzer} has the
advantage that any known microlensing event from the $\sim 100\,\rm deg^2$
that are currently monitored can be targeted.  In contrast to other
classes of interesting microlensing events, most particularly free-floating
planets, events containing massive remnants 
tend to be quite long, so that neither 
the low survey cadence in the majority of these fields nor the relatively
long time required to upload targets presents a serious obstacle.
A favorable feature of these binaries relative to other microlensing
binaries is that, due to their typically small microlens parallax
$\pi_\e=\sqrt{\pi_\rel/\kappa M}$, if the binary is seen to be in a caustic
trough from Earth, it is likely to also be in the trough as seen from 
{\it Spitzer}.  Hence, it is likely that at least one of
the sharp features induced by
caustic crossings will be monitored.  However, as discussed in 
Section~\ref{sec:break}, such events will generally not lead to
a precise mass measurement unless there are additional features
in the light curve, such as a post-caustic-exit bump due to a
cusp approach.  This implies that systems that are monitored
from before the caustic entrance, and therefore not usually known
to be binaries (e.g., OGLE-2015-BLG-1285) will be the most favorable
for precise mass measurements.

Such measurements for these systems would greatly benefit from more
aggressive observations from both the ground and {\it Spitzer}.  In the
present case, we were fortunate that the microlensed source was
a clump giant, with roughly 10 times larger radius than the Sun.
This meant that the caustic features lasted 10 times longer than for
a solar-type source.  Given the relatively sparse coverage over the
peak, it is possible that critical portions of 
these features would have been missed entirely
from the ground if the source had been similar to the Sun.  
The {\it Spitzer} observations were even sparser,
with only two over the bump.  There are a number of modifications that
could be made to more than double the cadence of {\it Spitzer} observations
(given the same overall time allocation).  In particular, with new
real-time reductions (Calchi Novati 2015, in prep), it should be possible
to stop observations of many events whose {\it Spitzer} light curve
has essentially reached baseline.  In addition, it is probably more productive
to increase overall cadences at the expense of the 2015 campaign's extra
observations for events that are relatively high magnification as seen
from the ground.

For OGLE-2015-BLG-1285, surveys were almost entirely responsible for
capturing the peak of the event from the ground (see Section~\ref{sec:obs}).
In this respect, two surveys that were specifically created for the
{\it Spitzer} campaign (UKIRT and Wise northern bulge surveys) as well
as the wide-area low-cadence survey undertaken by KMTNet in support
of this program, played an important role.  The continuation of such
surveys and the organization of new ones will be crucial.  In particular,
we note that the VISTA telescope is well placed to do such a survey
in highly extincted bulge regions.

However, it is also the case that followup observations could greatly
enhance the chances of detecting caustic features in areas that
are not well covered by surveys.  The primary motivation for such
followup observations to date has been planet detection.  This
both drives the allocation of available followup resources, which
are focused on planet sensitivity and detection, and also
fundamentally limits the total amount of followup resources to those
available to planet hunters.
For example, the very intensive LCOGT {\it Spitzer} support campaign
ran out of allocated observing time (due to ``too much'' good weather!)
two days before the peak of OGLE-2015-BLG-1285.  While the LCOGT
team did arrange to get some additional points of this event in
response to the anomaly alert, the main point here is that detection
and characterization of BHs is very challenging and observational
resources have been limited partly because the potential to detect
BHs is not widely appreciated.

The K2 microlensing campaign, scheduled for 83 days beginning April 2016
will provide a unique opportunity for space-based microlensing without
the need for ground-based alerts, and hence with a greater chance
that caustic entrances will be monitored from space.  This advantage
(relative to {\it Spitzer}) is balanced by the fact that these events
will be drawn from a relatively small area, albeit one with close
to a peak surface density of microlensing events.  From the standpoint
of making binary-lens mass measurements, and BH-binary mass measurements
in particular, we note that it is exceptionally important that this
entire area be monitored from the ground at high cadence and as continuously
as possible.  For example, even extremely faint stars can give rise
to briefly bright caustic crossings that can be effectively monitored
by {\it Kepler} with its 30 minute cadence, even if the majority of
the light curve cannot.  However, only if the corresponding caustic
crossings are monitored from the ground well enough to effectively model
the light curve, will this result in accurate mass measurements.
A very aggressive attitude toward continuous coverage will be especially
important toward the beginning of the campaign when individual southern
sites can observe the bulge for only five hours per night.

Gaia will provide complementary information on BH binaries.  There
are $\sim 3\times 10^5$ G dwarfs within 250 pc of the Sun, and 
it is hoped that these
will all have $\sigma(\pi)\sim 7\muas$ parallaxes by the end of the
5-yr mission.
For those that have BH companions with periods less than the mission
lifetime, the semi-major axes of their orbit about the
binary center of mass could be measured with the same precision.
This would imply $7\,\sigma$ detections for all those in the 
semi-major axis range $ 0.015\,\au< a< 5.3\,\au$ (assuming $M_{BH}=5\,M_\odot$).
Even the 1.5-year data release would enable detection of those in the
range $ 0.027\,\au< a< 2.4\,\au$.

Of course, Gaia cannot detect systems that, like
OGLE-2015-BLG-1285L, lie in the Galactic bulge, nor
can it detect totally dark systems to which microlensing is sensitive.
However, {\it WFIRST} will enable two different
probes of BHs that explore both of these regimes.  First, it will
be able to find and measure the mass of isolated BHs, as well as BH binaries
by combining microlens parallax measurements from its superb photometry
with astrometric microlensing from its excellent astrometry \citep{gouldyee14}.
For discussion on possible astrometric microlensing measurments with Gaia
see \citet{belokurov02}.
Second, it will obtain $\sigma(\pi)< 4\,\muas$ astrometry on of order 
$4\times 10^7$ 
stars in its microlensing fields \citep{gould15}.  The majority
of these stars will be in the Galactic bulge, where this precision
corresponds to a $7\,\sigma$ threshold of $\sim 0.25\,\au$.  Hence,
{\it WFIRST} will be sensitive to BH companions to $\sim 4\times 10^7$ 
luminous stars over the range $0.3\,\au<a<5.3\,\au$.

\acknowledgments 
Work by YS was supported by an
appointment to the NASA Postdoctoral Program at the Jet
Propulsion Laboratory, administered by Oak Ridge Associated
Universities through a contract with NASA.
The OGLE project has received funding from the National Science Centre,
Poland, grant MAESTRO 2014/14/A/ST9/00121 to AU.
Work by JCY, AG, and SC was supported by JPL grant 1500811.
Work by JCY was performed under contract with the California Institute of Technology 
(Caltech)/Jet Propulsion Laboratory (JPL) funded by NASA
through the Sagan Fellowship Program executed by the
NASA Exoplanet Science Institute.
Work  by  C.H.  was  supported  by  Creative  Research  Initiative
Program (2009-0081561) of National Research Foundation  of  Korea.
This research has made the telescopes
of KMTNet operated by the Korea Astronomy and Space Science
Institute (KASI).
L. W. acknowledges support from the Polish NCN Harmonia grant No.
2012/06/M/ST9/00172.
DM and AG acknowledge support by a grant from the US Israel Binational Science Foundation.
Work by DM is supported by the I-CORE program of the Israel Science Foundation and the Planning and Budgeting Committee.
This publication was made possible by NPRP grant \# X-019-1-006 from
the Qatar National Research Fund (a member of Qatar Foundation)
S.D. is supported by “the Strategic Priority Research Program—The
Emergence of Cosmological Structures” of the Chinese Academy of
Sciences (grant No. XDB09000000).
Work by SM has been supported by the Strategic Priority Research
Program ``The Emergence of Cosmological Structures" of the Chinese
Academy of Sciences Grant No. XDB09000000,
and by the National Natural Science Foundation of
China (NSFC) under grant numbers 11333003 and 11390372.
M.P.G.H. acknowledges support from the Villum Foundation. Based on
data collected by MiNDSTEp with the Danish 1.54 m telescope at the ESO
La Silla observatory.
This work is based in part
on observations made with the {\it Spitzer} Space Telescope,
which is operated by the Jet Propulsion Laboratory, California Institute of Technology under a contract with
NASA.
The United Kingdom Infrared Telescope (UKIRT) is supported by NASA and
operated under an agreement among the University of Hawaii, the University
of Arizona, and Lockheed Martin Advanced Technology Center; operations are
enabled through the cooperation of the Joint Astronomy Centre of the Science
and Technology Facilities Council of the U.K.
This work makes use of observations from the LCOGT network, which
includes three SUPAscopes owned by the University of St Andrews. The
RoboNet programme is an LCOGT Key Project using time allocations from
the University of St Andrews, LCOGT and the University of Heidelberg
together with time on the Liverpool Telescope through the Science and
Technology Facilities Council (STFC), UK. This research has made use
of the LCOGT Archive, which is operated by the California Institute of
Technology, under contract with the Las Cumbres Observatory.
{\bf The \textit{Spitzer} Team thanks Christopher S.\ Kochanek for graciously trading us his allocated observing 
time on the CTIO 1.3m during the \textit{Spitzer} campaign.}

Copyright 2015. All rights reserved.


\begin{table}
\centering
\caption{Best-fit microlensing model parameters
and their 68\% uncertainty range derived from the MCMC chain density
(for both $u_0>0$ and $u_0<0$).
We note there are non-linear correlations among the parameters,
and that the entire $\chi^2$ surface is not parabolic.
The solutions are almost symmetric, with a small difference in $\pi_{\rm E}$,
due to the small offset between the projected $Spitzer$-Earth axis and the ecliptic.
\label{tab:model}}
\begin{tabular}{lll}
\\
\tableline\tableline
Parameter & $u_0>0$ & $u_0<0$\\
\tableline
$t_0-2457200$ [d] & 9.74 [9.73,9.75] & 9.74 [9.73,9.75]\\
$ u_0$ & 0.46 [0.42,0.53] & -0.46 [-0.42,-0.53]\\
$t_{\rm E}$ [d] & 31.4 [29.5,32.5] & 31.4 [29.5,32.5]\\
$\pi_{\rm E,N}$  & 0.019 [0.012,0.029] & -0.018 [-0.012,-0.029]\\
$\pi_{\rm E,E}$  & 0.0087 [0.0075,0.0112] & 0.0092 [0.0080,0.0121]\\
$t_*$ [d]  & 0.455 [0.450,0.459] & 0.455 [0.450,0.459]\\
$q$ & 2.9 [2.1,3.4] & 2.9 [2.1,3.4]\\
$s$ & 1.934 [1.928,1.944] & 1.934 [1.928,1.944]\\
$\alpha$ [deg] & 80 [78,82] & -80 [-78,-82]\\
\tableline\tableline
\end{tabular}
\end{table}

\begin{table}
\centering
\caption{Physical properties of the binary system.
Median and 68\% uncertainty range values derived from the MCMC chain density,
after accounting for the Jacobian of the transformation between the MCMC variables
and the physical quantities.
\label{tab:physical}}
\begin{tabular}{lcc}
\\
\tableline\tableline
Parameter & Median & 68\% confidence intervals \\
\tableline
$M_1$ [$M_\odot$] & 2.0 & [1.2,3.3] \\
$M_2$ [$M_\odot$] & 0.8 & [0.5,1.2] \\
$r_\perp$\,\, [$\au$] & 6.1 & [5.7,6.5] \\
$D_L$ [kpc] & 7.5 & [7.3,7.7] \\
\tableline\tableline
\end{tabular}
\end{table}

\begin{figure}
\plotone{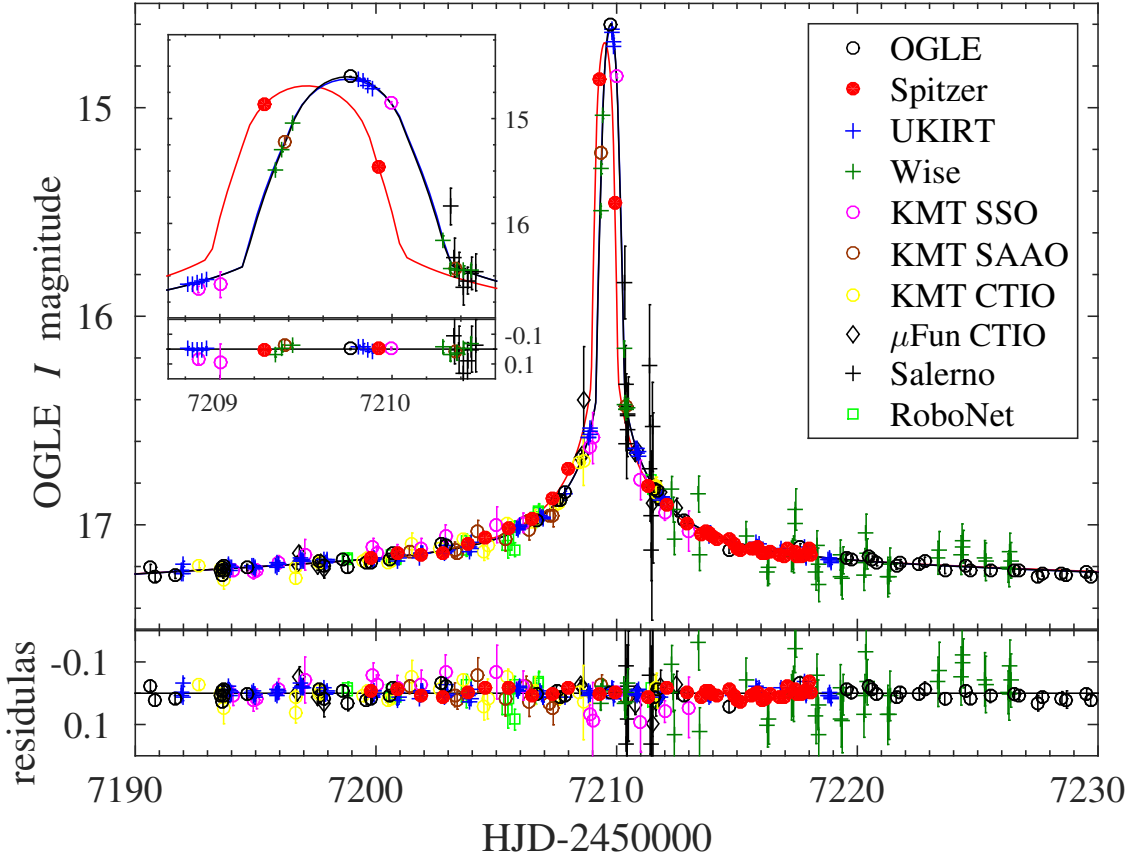}
\caption{Light curve of OGLE-2015-BLG-1285 with data from {\it Spitzer} (red)
and various ground-based observatories (see interior figure labels).
The magnitude scale is what is directly observed by OGLE.  All other
observatories (including {\it Spitzer}) are aligned so that equal ``magnitude''
reflects equal magnification.  The very small $(\sim 0.3\,\rm day)$ offset
between the peak as seen by {\it Spitzer} and the ground hints that the
microlens parallax $\pi_\e=\sqrt{\pi_\rel/\kappa M}$ may be small, which
would imply a high-mass lens.  For ground-based data, 
two models are shown, one for $H$-band limb darkening (blue),
which should be compared to UKIRT data, and one for $I$-band (black), which
should be compared to all other data.
The difference between those two curves can be seen only at the peak of the anomaly.
}
\label{fig:lc}
\end{figure}

\begin{figure}
\plotone{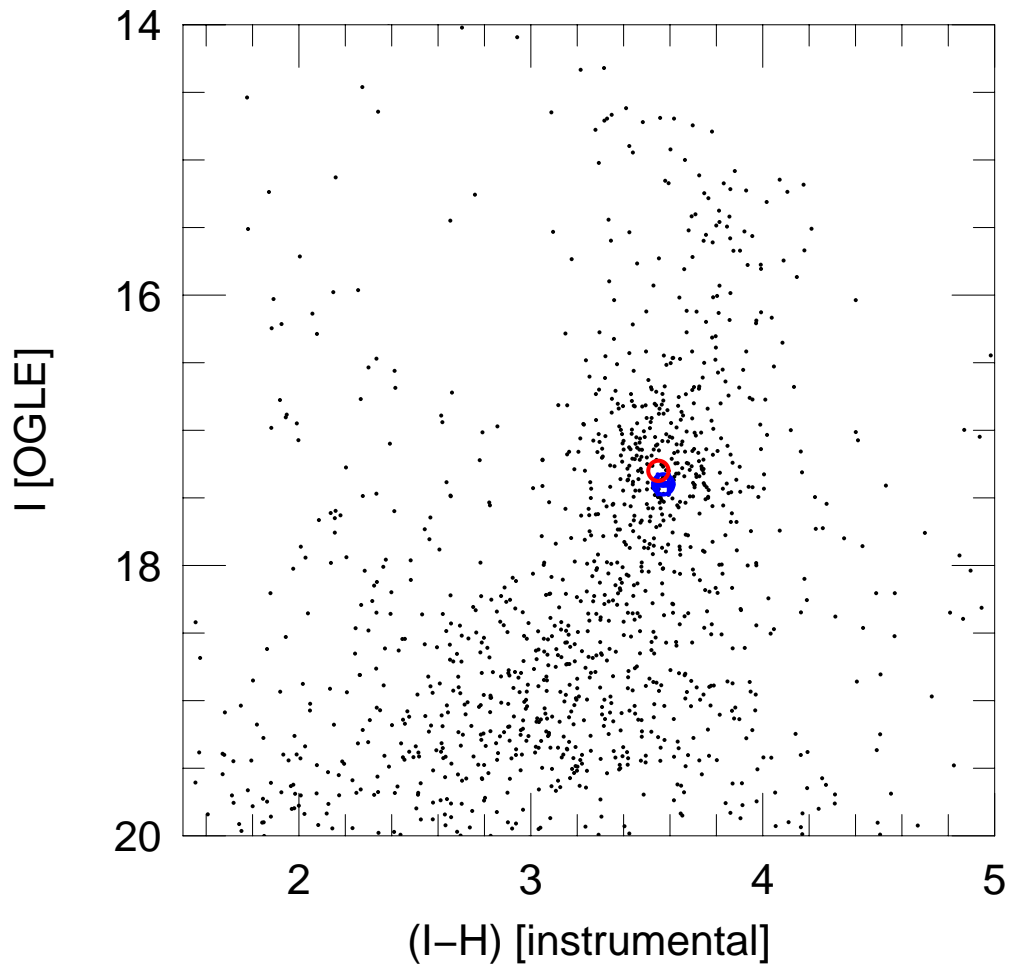}
\caption{CMD derived by combining calibrated $I$-band photometry from OGLE
with instrumental $H$-band photometry from UKIRT.  The centroid of the
clump (red) and the ``baseline object'' (blue) are marked. From this offset one
derives the source radius $\theta_* = 6.01\,\muas(f_s/f_{\rm base})^{1/2}$,
where $f_s/f_{\rm base}$ is the ratio of source flux to baseline flux.
}
\label{fig:cmd}
\end{figure}

\begin{figure}
\plotone{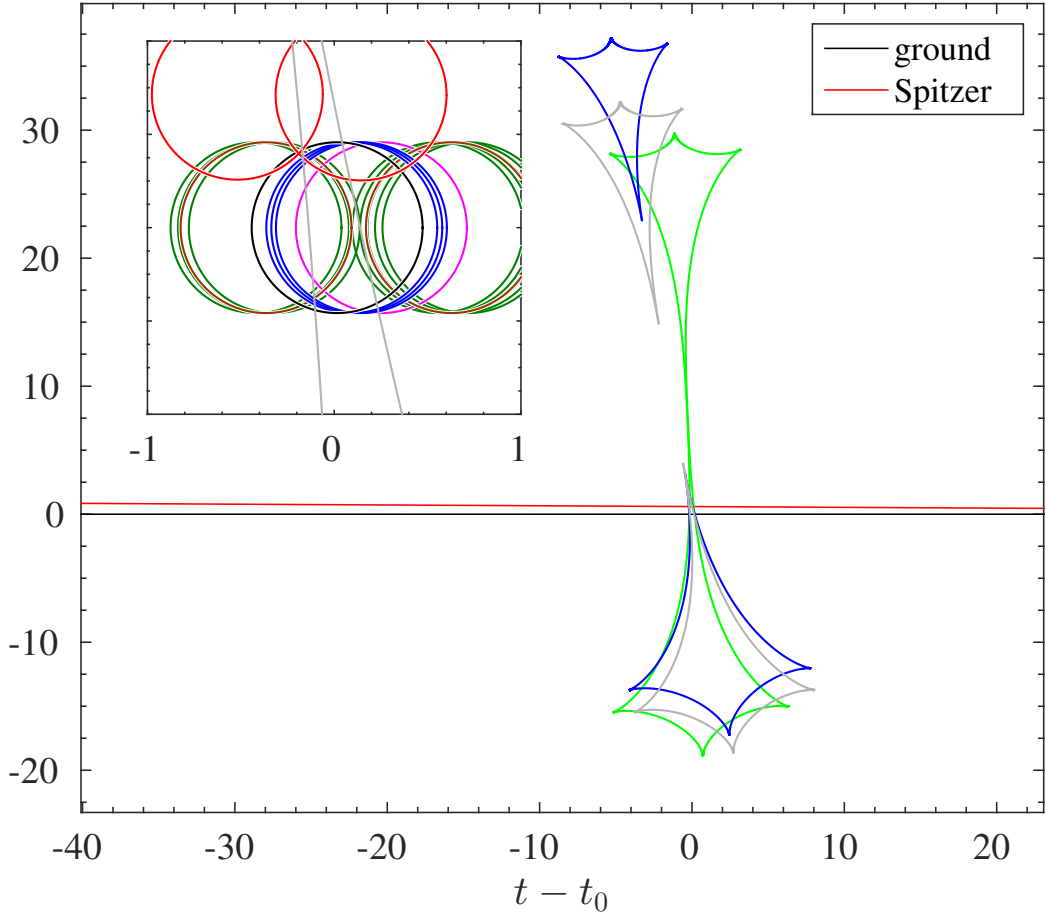}
\caption{Three different caustic structures that
are consistent with the light curve (Fig.~\ref{fig:lc}), i.e.,
from the $2\,\sigma$ region of Figure~\ref{fig:inner}.  All three
are rotated by $\alpha$ and scaled by $t_\e$ so that the x-axis is simply
source position as a function of time.  Hence, the binary axis is oriented
so that the primary is toward the top of the plot and the secondary
is toward the bottom.  Both wide-binary and resonant
caustic topologies are permitted.  Main panel shows full caustics with
{\it Spitzer} trajectory shown for one of the three caustic structures.
Inset is a zoom showing the source size and its position at times
of observations, with same color scheme as Figure~\ref{fig:lc},
in which {\it Spitzer} points are shown relative to the caustic
rather than clock time.  An alternate topology, which is ruled out
by arguments given in Sections~\ref{sec:gblc} and \ref{sec:astrochrom}, would
have the source trajectory pass through the bottom-most cusp, nearly
perpendicular to the binary axis (not shown).
}
\label{fig:caust}
\end{figure}

\begin{figure}
\plotone{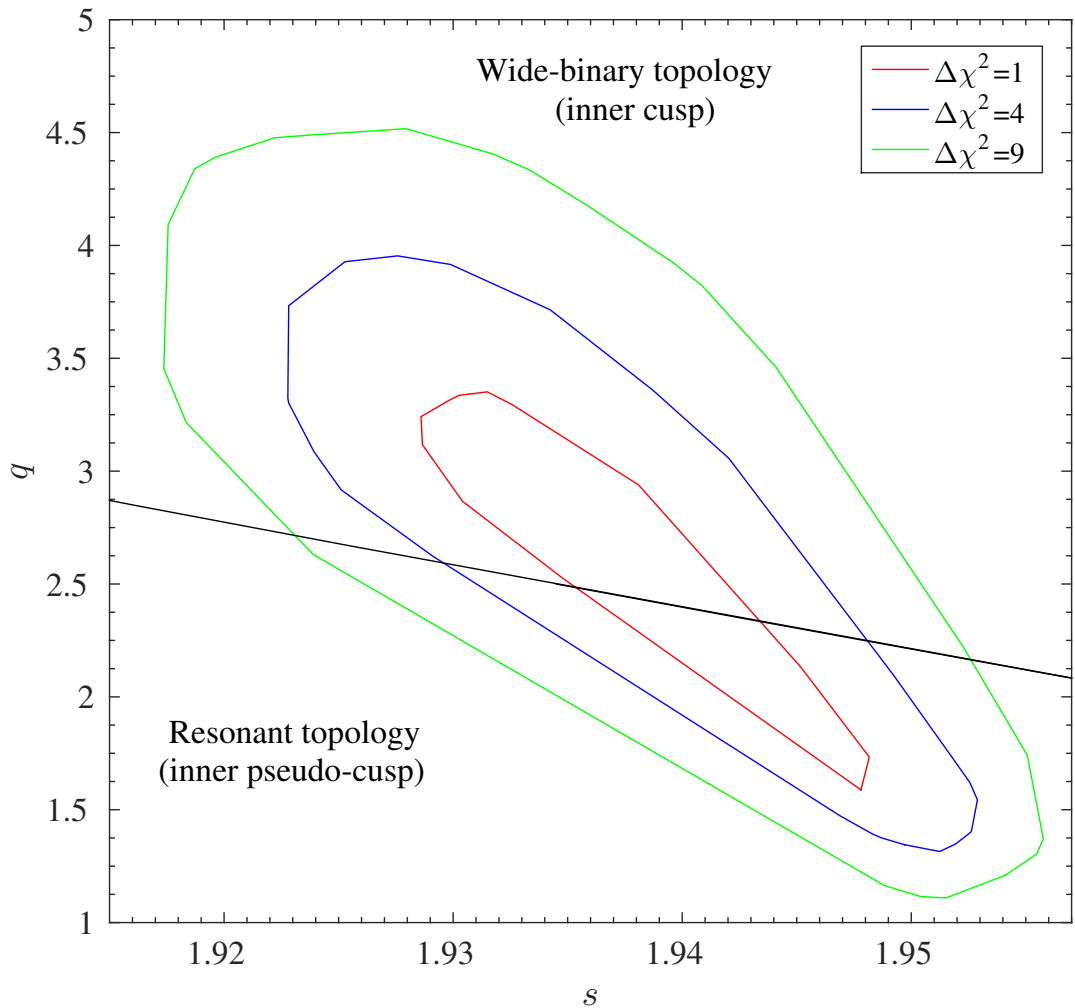}
\caption{$\chi^2(s,q)$ surface for ``inner cusp'' topology, where
$s$ is the projected separation of the components normalized to $\theta_\e$
and $q$ is the mass ratio of the primary to its companion.  Black curve
$s^2 = (1 + q^{1/3})^3/(1+q)$ shows boundary between wide-binary topology
(two 4-sided caustics) and resonant topology (one 6-sided caustic).
See Figure~\ref{fig:caust}.
}
\label{fig:inner}
\end{figure}

\begin{figure}
\plotone{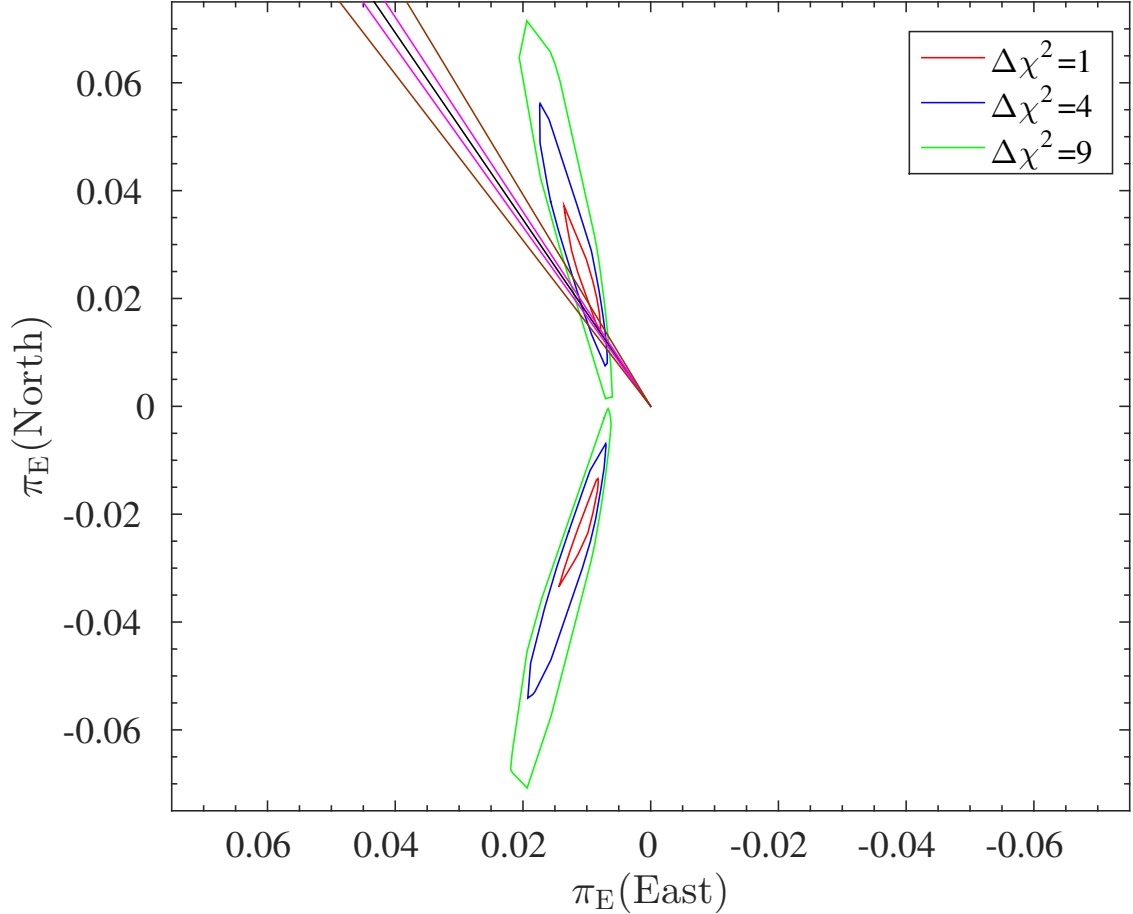}
\caption{$\chi^2(\bpi_\e)$ surface.  Solutions
with $u_0>0$ ($u_0<0$) lie in the upper (lower) part of the diagram.
Physical explanation for quasi-1-D contours is given by 
Equations~(\ref{eqn:pieconstraint}) and (\ref{eqn:pieconstraint2}) in
Section~\ref{sec:parms+pie}.  \bf{The magenta and brown rays converging at the origin
show the impacts for hypothetical future proper-motion measurements
$\hat{\mu}=(60\pm 1)^\circ$ and $\hat{\mu}=(60\pm 3)^\circ$,
respectively.}
}
\label{fig:pie}
\end{figure}

\begin{figure}
\plotone{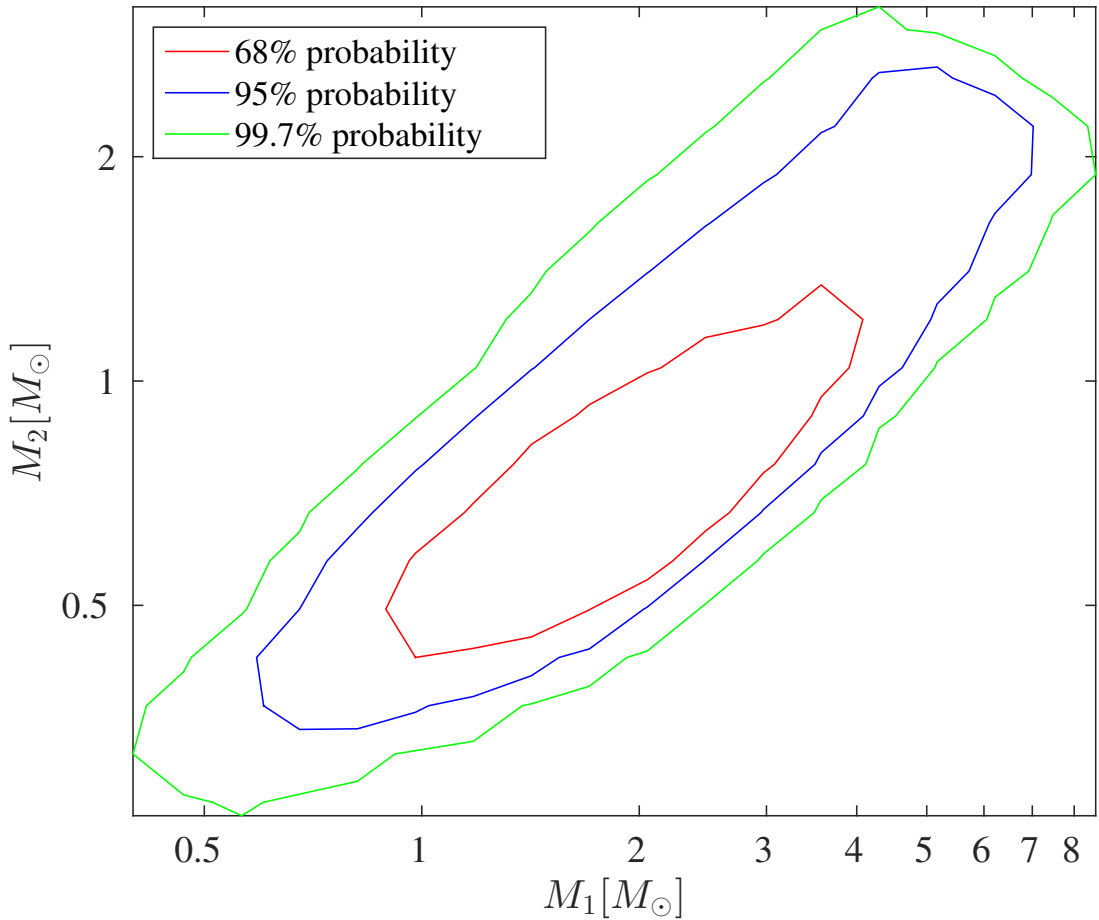}
\caption{Binary mass map. Contours showing 68\%, 95\% and 99.7\% probability regions.
The most probable combination is
a NS primary and a main-sequence secondary.  However, binaries
with two main-sequence stars or two massive remnants 
(e.g., BH+NS) are also possible.
}
\label{fig:mass}
\end{figure}

\begin{figure}
\plotone{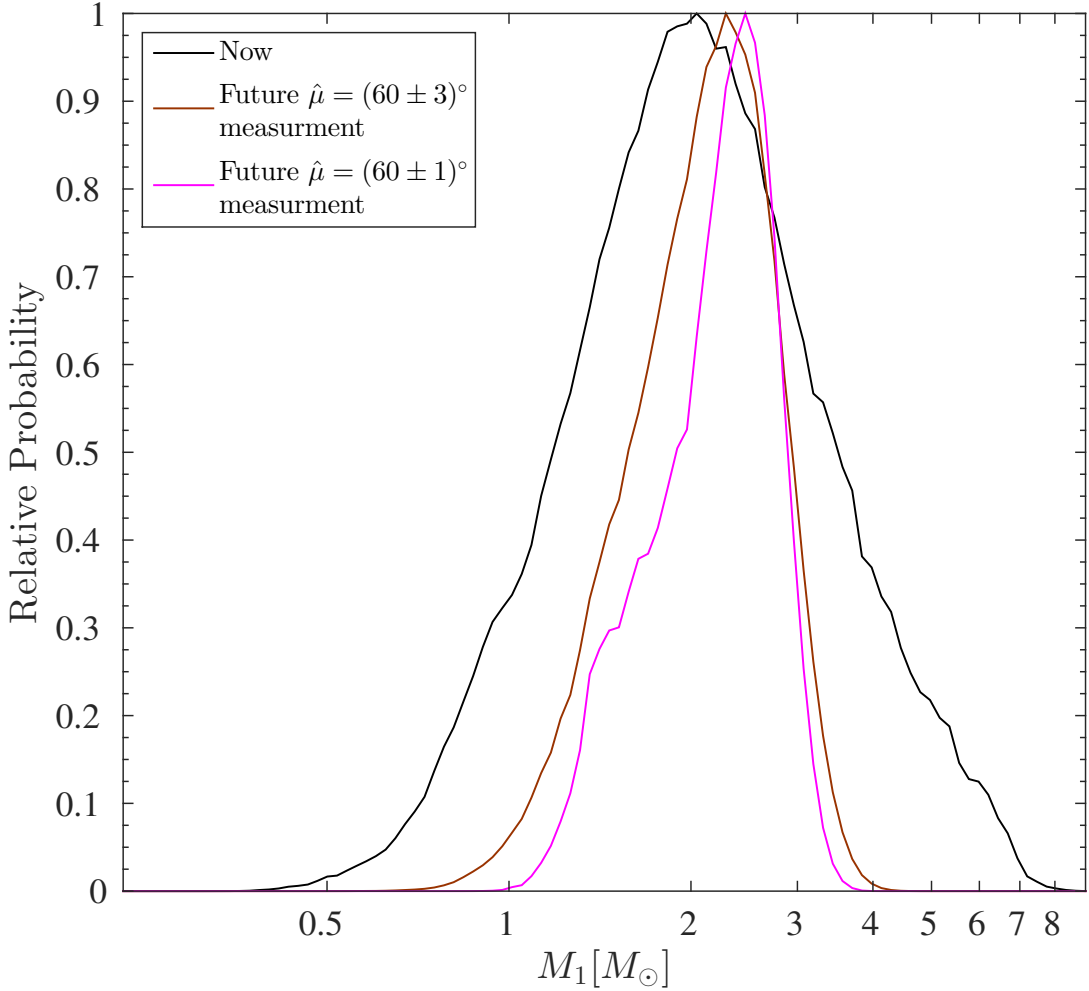}
\caption{Differential probability distribution of the primary mass $M_1$,
assuming a prior uniform in $\log(M)$,
as of ``now'' (black), i.e., based solely on microlensing measurement.
About 80\% of the probability lies above $M_1>1.35\,M_\odot$,
making this a massive-remnant (NS or BH) candidate.  Magenta and brown
curves show the impact of future
proper motion measurements $\hat{\mu}=(60\pm 1)^\circ$ and $\hat{\mu}=(60\pm 3)^\circ$,
respectively, as indicated in Figure~\ref{fig:pie}.
}
\label{fig:diffm}
\end{figure}

\end{document}